\documentclass[letterpaper,twocolumn,10pt]{article}

\usepackage[latin1]{inputenc}
\usepackage[english]{babel}
\usepackage{scs}
\usepackage{graphicx}
\usepackage{amssymb,amsmath}
\usepackage{color,listings}

\definecolor{light-gray}{gray}{0.9}
\begin{document}

\title{Multilateration: Methods For Clustering Intersection Points \\
For Wireless Sensor Networks Localization With Distance Estimation Error\footnotemark[1]
}

\author{
Marios Karagiannis~\footnotemark[2], Ioannis Chatzigiannakis~\footnotemark[3] and Jose Rolim~\footnotemark[2] \\
\footnotemark[2] {\rm Centre Universitaire d' Informatique, Geneva, Switzerland. 
{\small E-mail: {\tt \{marios.karagiannis,jose.rolim\}@unige.ch}}}\\
\footnotemark[3] {\rm Research Academic Computer Technology Institute (CTI) and University of Patras, Greece. {\small E-mail: {\tt ichatz@cti.gr}}}
}
\maketitle

\footnotetext[1]{This work has been partially supported by the IST Programme of the European Union under contract number ICT-2008-224460 (\textsf{WISEBED})}


\begin{abstract}
\noindent
In this paper we describe three methods for localizing a wireless sensor network node, using anchor nodes in its neighbourhood, when there is an error in distance estimation present. We use the intersection points of the circles formed with the estimated distances from each anchors and we apply different methods to form clusters. We then use the cluster points to calculate the final position.
\end{abstract}

\section{Introduction}
\label{sec:intro}

A particularly promising and hot research area has been the design and analysis of \textit{wireless sensor networks} (WSN), which has attracted researchers from very different backgrounds, such as hardware, software, algorithms, and data structures, as well as researchers from various application areas. Development and evaluation of dependable wireless sensor systems requires answering many design questions. Advancements have been made in the physical hardware level, embedded software in the sensor devices, systems for future sensing applications and fundamental research in new communication and networking paradigms.  

One of the major issues that wireless sensor networks deployment have, is the issue of localization. Localization is the parallel process of calculating the position of each of the nodes that belong to the network. This is accomplished usually using a few nodes with either predefined positions or special hardware that enable them to know their positions beforehand. These "special" nodes are usually called Anchors or Beacons.

\subsection{Localization in Wireless Sensor Networks}
\label{sec:wsnlocalization}
Localization is an important research area in wireless sensor networks. It provides the nodes that comprise the network the ability to know their location in the area the network is deployed, which in turn allows the network itself to perform tasks that would be impossible to perform without this knowledge. Applications for wireless sensor networks that demand the presence of location information include wildfire detection, target tracking and battlefield observation networks. Each of the above list requires the network to perform tasks that are based on physical locations within the network. The feedback from the network will also include position information so an initialization step is critical before these can function properly.
The location of a single node in a wireless sensor network can be obtained using several methods. The simplest of them is using a GPS chip, which have become significantly smaller, more energy efficient and less expensive than in the past, to a point that including one is no longer a costly decision, both in terms of energy consumption and of actual cost of the device. The GPS system provides location information using a network of satellites that orbit the earth, so even a single node with no neighbouring nodes can be localized. Traditionally, only a few nodes in a network, called anchors or beacons, were equipped with a GPS chip, and several methods were used to take advantage of their position information to localize the rest of the network. Some of these techniques can be found in \cite{local1},\cite{local2},\cite{local3}.Although now we are in a position to equip many more nodes with similar technology, these techniques are far from obsolete. The reason is that the GPS has a few drawbacks that make its use less than ideal in certain circumstances. Two examples can be use of wireless sensor networks in other planets, losing the advantage of having the satellite network that provides the GPS information installed and indoors where the satellite signal is weak at best and non existent typically. In these cases, among others, localization techniques that use anchor nodes are very much relevant.
Localization techniques can provide a node with either an approximation of its absolute position or even coordinates in a new system, which is useful only this particular network \cite{vcs1}\cite{vcs2}\cite{vcs3}. In the first case, the goal is of course to provide an approximation as close as possible to the real location. The assessment of such algorithms counts exactly that, the error between the approximated location and the real location of the nodes.

\section{Related work}
\label{sec:related}

The are of Wireless Sensor Networks localization has been investigated for many years. The localization process usually requires a number of special nodes to already know their location. In range based algorithms, these special nodes, often called anchors, advertise their locations and nodes that wish to be localized use this information along with an estimation of their distance to the anchors to calculate their positions. 
Range based algorithms include the ones in \cite{Savvides01dynamicfine-grained}, \cite {Savarese01robustpositioning}, \cite{AspnesEGMWYAB2006} and use the above technique with newly localized nodes becoming anchors in the next step of the algorithm. Techniques like these produce an increased number of transmissions, which in turn result in increased power usage in the network.
In range free algorithms, such as the one in \cite{Bulusu00gps-lesslow} and \cite{Niculescu03adhoc} the overhead is even larger due to increased communication required for the algorithms to function, without the ability to measure distances from each other.

\subsection{Error in distance estimation}
\label{sec:errorindistance}
Error in distance estimation between wireless nodes in inevitable regardless of the method used to calculate the distance.
In \cite{Savvides01dynamicfine-grained} a few techniques were tested, including RF and Ultrasound Time of Arrival and received signal strength indicator (RSSI). The concluded that some techniques such as ToA are better suited for localization than others, although RSSI is a cheaper, simpler in hardware demands and more widely available. They also note that using the RSSI gives results that vary greatly depending on the conditions of the experiment. As we will see, similar results were given by other researchers regarding the RSSI usage.
Using the RSSI is a widely used method of estimating the distance between two nodes. The RSSI is often provided automatically by the hardware of the nodes. Studies such as \cite{1298094} have shown that factors such as the radio frequency, transceivers variations, antenna orientation, positions of the nodes including elevation play a significant role on the estimated distance calculated from the RSSI. Given some constraints in the above factors, coming up with a model that resembles reality is feasible but a generic model is very hard to provide.
As shown in \cite{1234829} using RSSI for ranging-based localization is feasible alternative to GPS. They achieved a 4.1m error in a 49 node network deployed in a half-football field sized area. They also noted that the conditions of the experiments play a significant role in this error, since the above error estimation can fluctuate depending on things like nodes elevation or grass height.
Error-free distance calculation between anchor nodes and nodes that require localization would provide an instant and trivial solution to the problem, but these calculations are often not possible. Different techniques can be used to get a better or worse estimation, but errors are still present.

\section{Motivation \& Our Contribution}
\label{sec:contrib}

\begin{figure}
\centering
\includegraphics[width=0.45\textwidth]{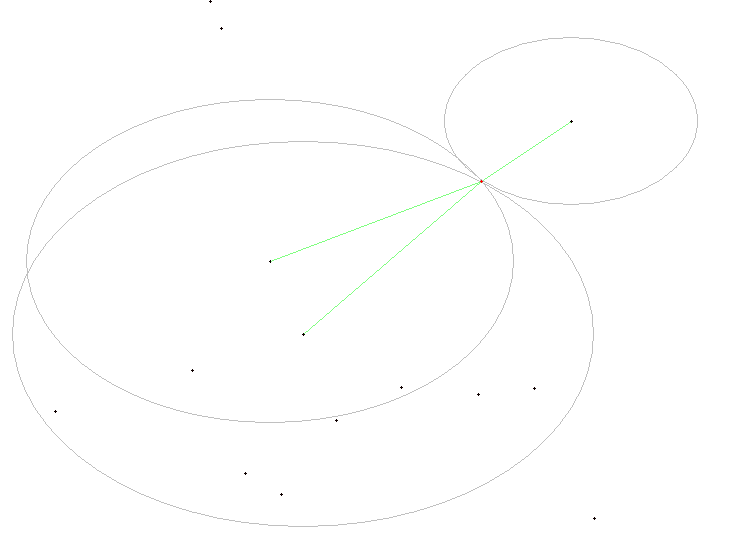}
\caption{\label{fig-1} Intersection Points Converging on Real Position}
\end{figure}

\begin{figure}
\centering
\includegraphics[width=0.45\textwidth]{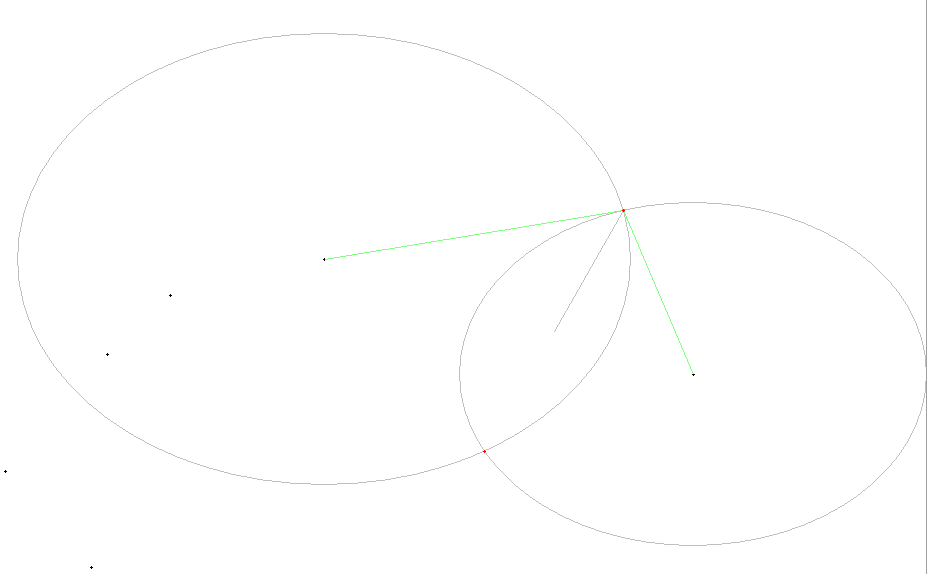}
\caption{\label{fig-2} Two Anchors Intersection Points}
\end{figure}

In this paper we try to explore different techniques that can be used to successfully localize each node in a wireless sensor network. For us, successful localization means that the computed final position of each node will be as close as possible to the real position. 
The technique we use takes into account the estimated distance of each node to its neighbour anchor nodes. We assume that each node has the means to determining this distance (for example, using RSS signal strength \cite{1234829}, delay of travel, optical, ultrasonic\cite{1127825}  or any other technique) with a percentage of error which can be due to different reasons such as the ones presented in \ref{sec:errorindistance}. The estimated distance for each anchor forms a circle, so for n anchors we draw n circles. In the special and ideal case where the error estimation is 0, half of the intersection points of all these circles converge to the real position of the node, as shown in Figure~\ref{fig-1}. This is true for $n>2$, because for $n=2$ we may have two different intersection points (Figure~\ref{fig-2}) and of course for $n=1$ we have no intersection points.

\begin{figure}
\centering
\includegraphics[width=0.45\textwidth]{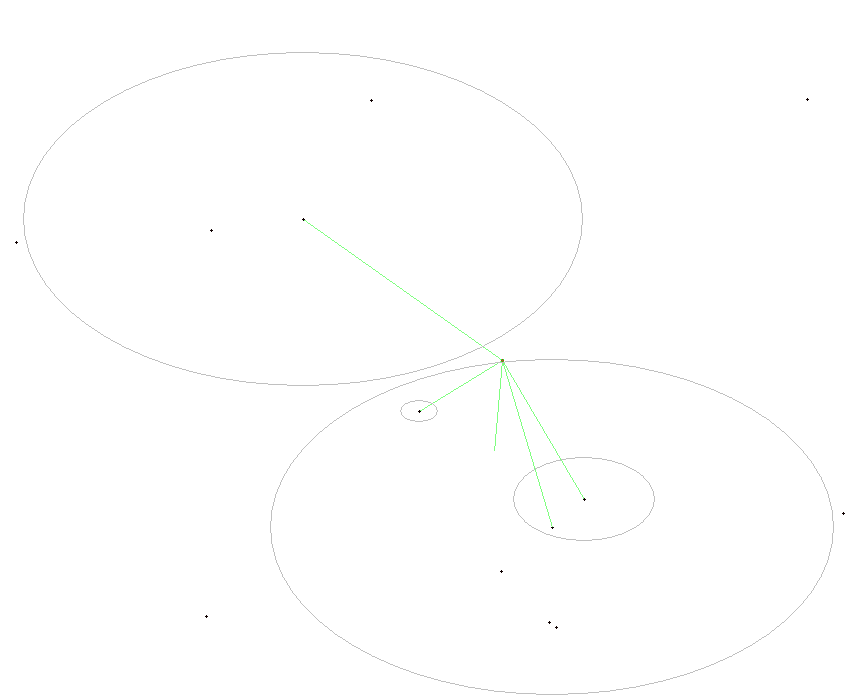}
\caption{\label{fig-3} Example of No Intersection Points}
\end{figure}

The number of intersection points for the n circles is greatly affected by the error percentage in distance estimation. In special cases, with relatively large errors there can be no intersection points Figure ~\ref{fig-3}. In this case trying to cluster the intersection points to calculate the position of a node is, of course, impossible. In our simulations, we set the error percentage as the maximum error percentage and we calulcated the error as a random percentage around the real distance in each iteration. We then did multiple iterations, in order to avoid the problem of no intersection points and to be able to calculate a position every time. We did this in order to be able to evaluate the accuracy of our methods in all cases.

\subsection{Error models}
\label{sec:errormodels}
We know that the signal strength decreases when the sender-receiver distance increases. If $EMax$ is the maximum power used to send a message from the sender and $ERec$ is the received power at the receiver, we call the relative decrease $Emax-ERec$ attenuation. The attenuation in theory, without obstacles and reflections, is proportional to the square of the distance ($att=d^2$), but in real experiments it is less than that. In order to model the errors in distance estimation, we considered 4 different error models. 
The first model is the Constant Error model. In this model, the estimated distance, compared to the real distance is given by the following formula:$EstDist=RealDist + e*MaxRange$. 

\begin{figure}
\centering
\includegraphics[width=0.45\textwidth]{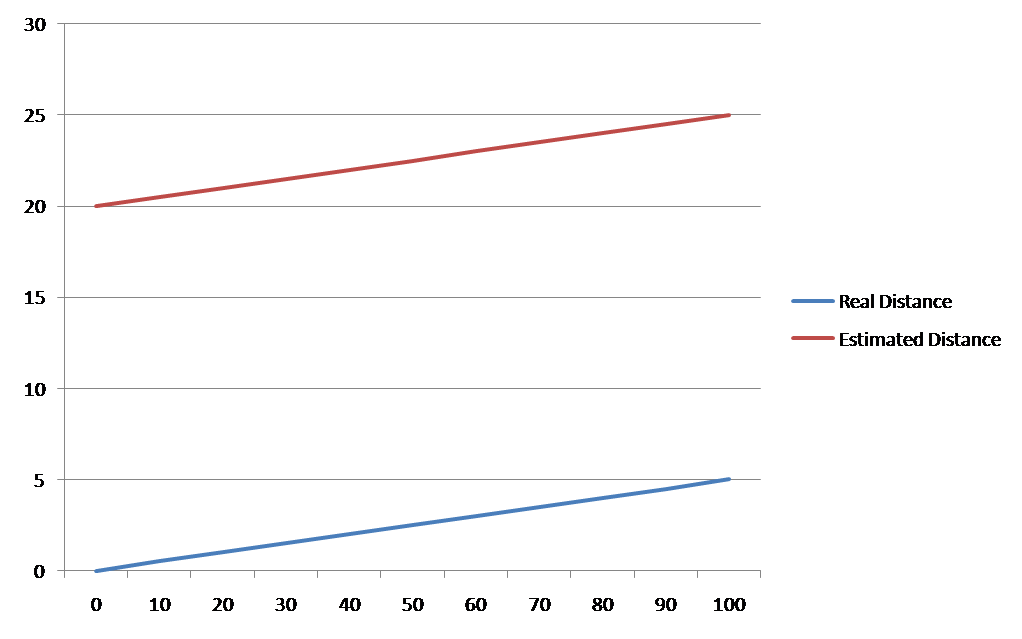}
\caption{\label{example_constant} Example of Constant Error model}
\end{figure}

\begin{figure}
\centering
\includegraphics[width=0.45\textwidth]{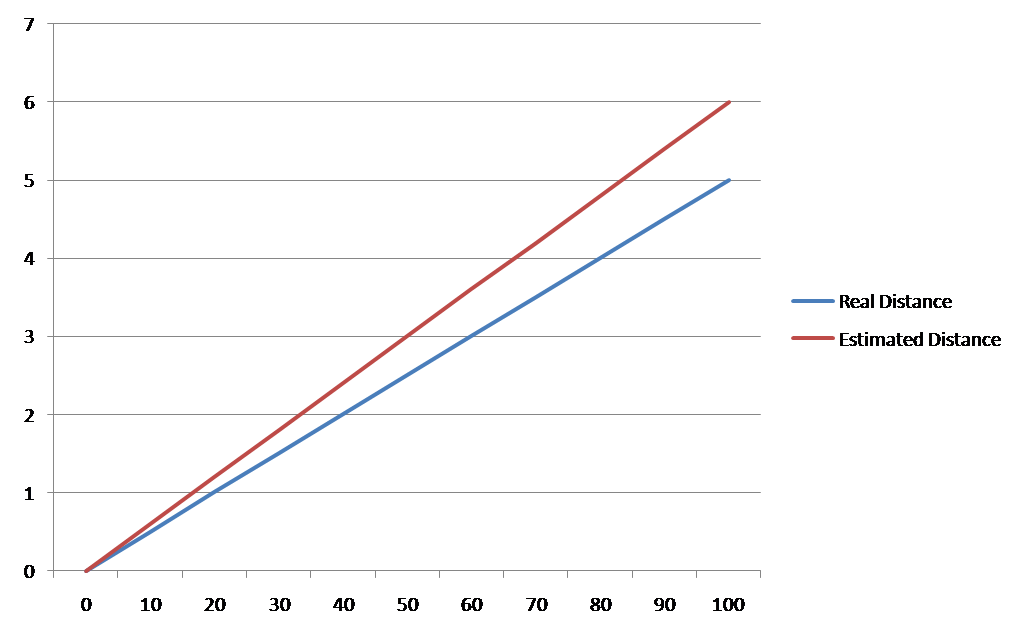}
\caption{\label{example_linear} Example of Linear Error model}
\end{figure}

\begin{figure}
\centering
\includegraphics[width=0.45\textwidth]{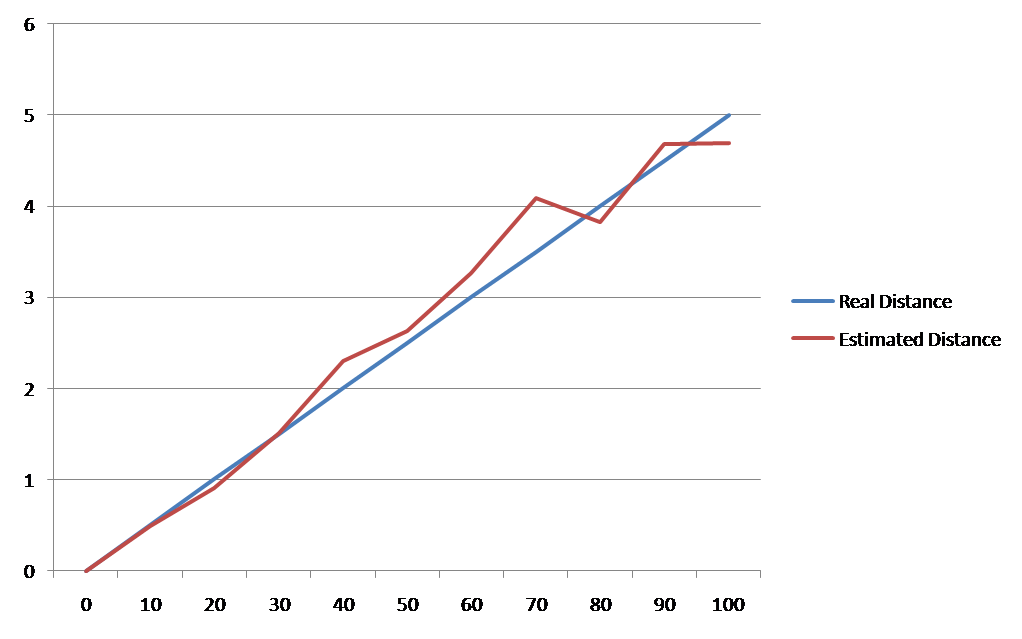}
\caption{\label{example_random} Example of Random Error model}
\end{figure}

\begin{figure}
\centering
\includegraphics[width=0.45\textwidth]{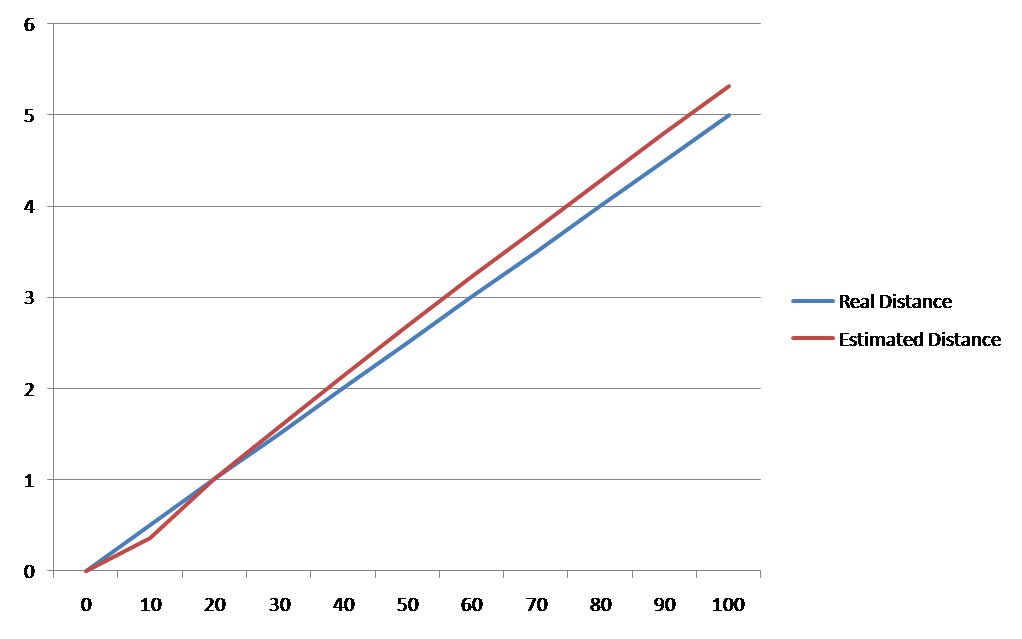}
\caption{\label{example_logarithmic} Example of Logarithmic Error model}
\end{figure}

The second model is the Random Error model. In this model, the estimated distance is given by the formula $EstDist=RealDist \pm random(e)*RealDist$. 
The third model is the Linear Error model. In this model, the estimated distance is given by the formula $EstDist=RealDist \pm e*RealDist$. 
The fourth model is the Logarithmic Error model. In this model, the estimated distance is given by the formula $EstDist=RealDist + a*e$ where $a=0$ when $distance=0$ or $a=ln(RealDist)*e$ when $distance>0$.
In all the above formulas $e$ is a percentage parameter where $0<e<1$. It is obvious that when $e=0$, the formulas become $EstDist=RealDist$ in all four models. We assume that $0 \leq RealDist \leq MaxDist$.
In figures  ~\ref{example_constant}, ~\ref{example_linear}, ~\ref{example_random}, ~\ref{example_logarithmic} we show examples of the 4 error models described above, using $e=0.2$ and $MaxDist=6$.

The four models described above are simple mathematical constructions that do not take into account multipath effects or environmental changes.

We used the data we acquired in \cite{TR-2010-UNIGE-RACTI-LQI} in order to evaluate the above models. In this technical report, an experimental test bed that is comprised by 6 "gathering stations" is used to get RSSI (LQI) values from a mobile node that moves between 82 fixed points around the gathering stations. For each position, 200 messages are being sent to all 6 gathering stations. The gathering stations report these values to a base station which then finds the average RSSI value for each location and each gathering station.
In order to get an idea of what an actual distance estimation based on RSSI values looks like, we took data from 4 straight lines that passed through one of the central gathering stations mentioned in the Technical report, in a star shape, and drew the RSSI values. 

We used a widely used propagation model, the log-normal shadowing model, which is described in \cite{1234829} to transform the RSSI values to distance values. In this model, the multipath effects can be taken into account, and the calculation of the is based on the following formula:

$RSSI(d)[dBm]=RSSI_0-10_nlog_{10}(\frac{d}{d_0})+X_\sigma$

where $n$ is an attenuation constant, $d$ is the distance, $X_\sigma$ is a zero-mean Gaussian with standard deviation $\sigma$ used to simulate the multipath effects and $RSSI_0$ is the signal strength at a reference distance $d_0$. In order to simplify the calculation of the distance, we can omit the multipath effects factor ($X_\sigma=0$) and use the following formula:

$d(RSSI)=10\frac{RSSI_0-RSSI}{10n}$

\begin{figure}
\centering
\includegraphics[width=0.45\textwidth]{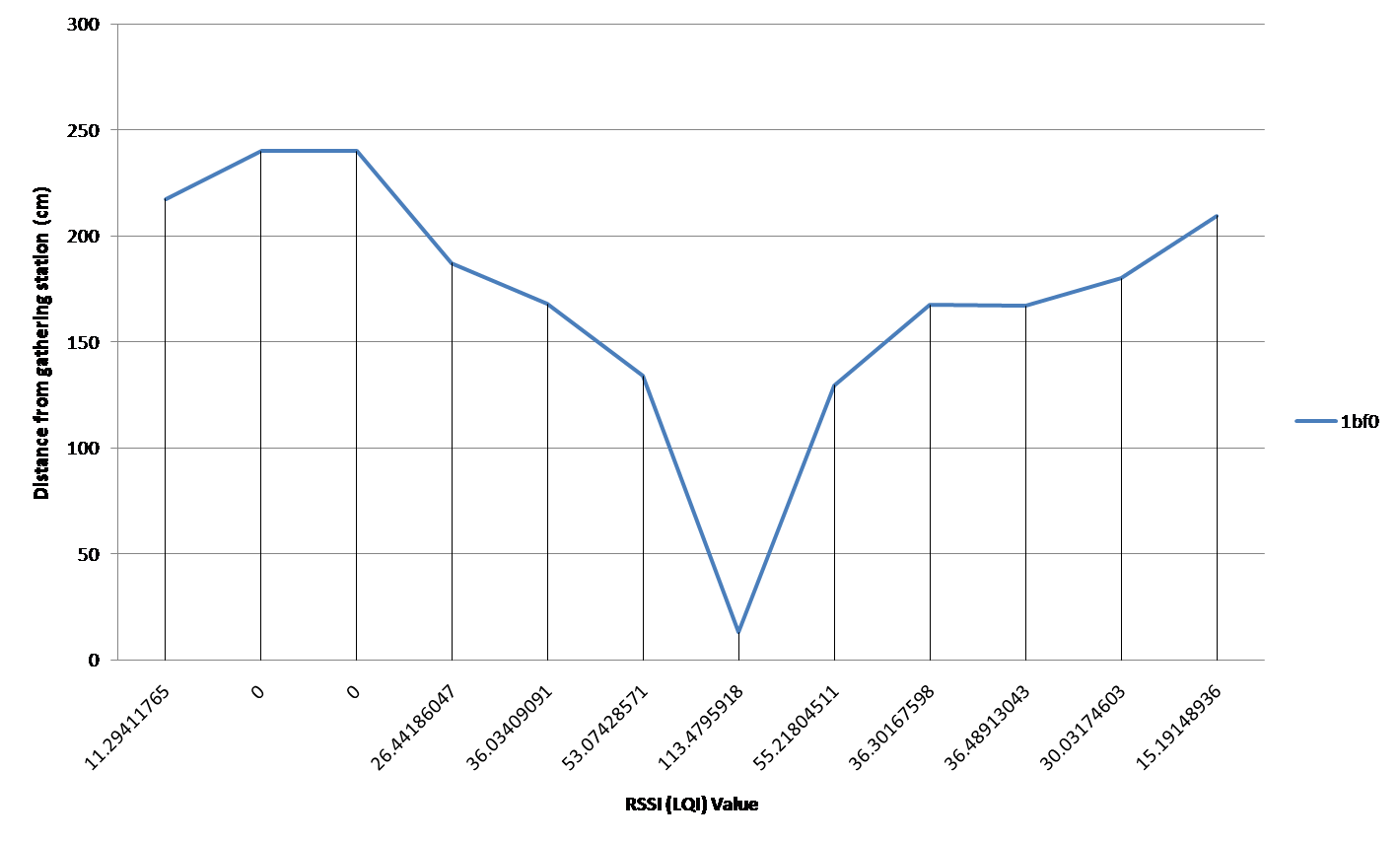}
\caption{\label{lqi1} Distance derived from experimental RSSI values}
\end{figure}

\begin{figure}
\centering
\includegraphics[width=0.45\textwidth]{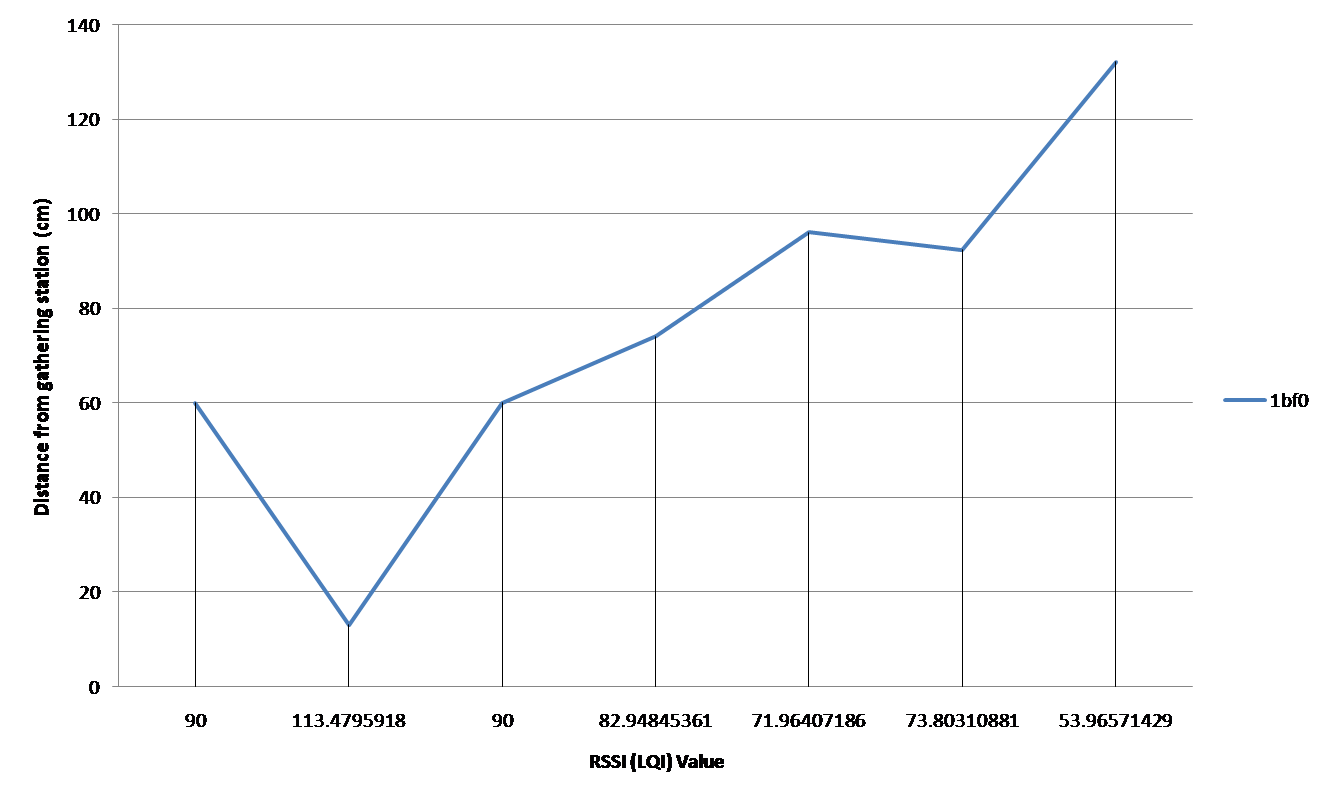}
\caption{\label{lqi2} Distance derived from experimental RSSI values}
\end{figure}

\begin{figure}
\centering
\includegraphics[width=0.45\textwidth]{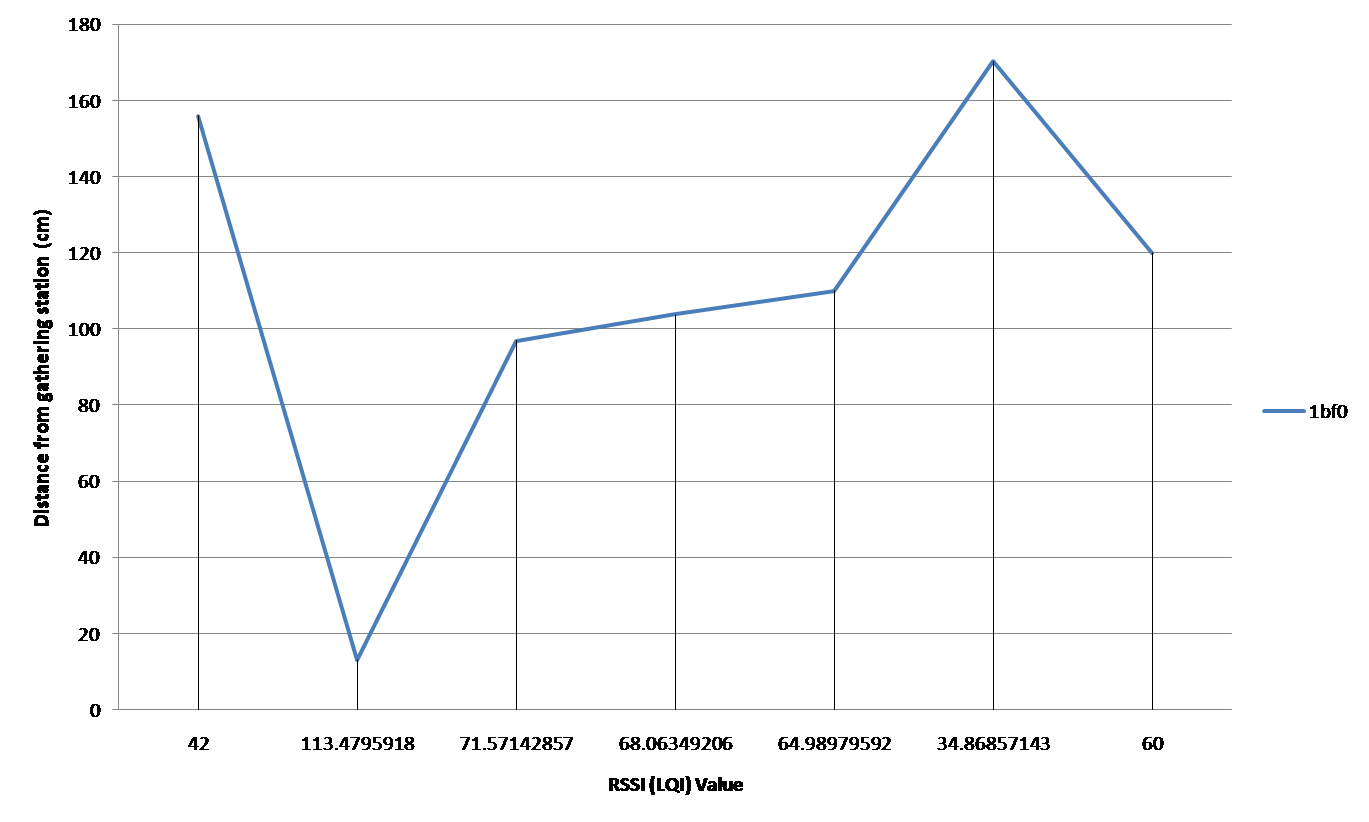}
\caption{\label{lqi3} Distance derived from experimental RSSI values}
\end{figure}

\begin{figure}
\centering
\includegraphics[width=0.45\textwidth]{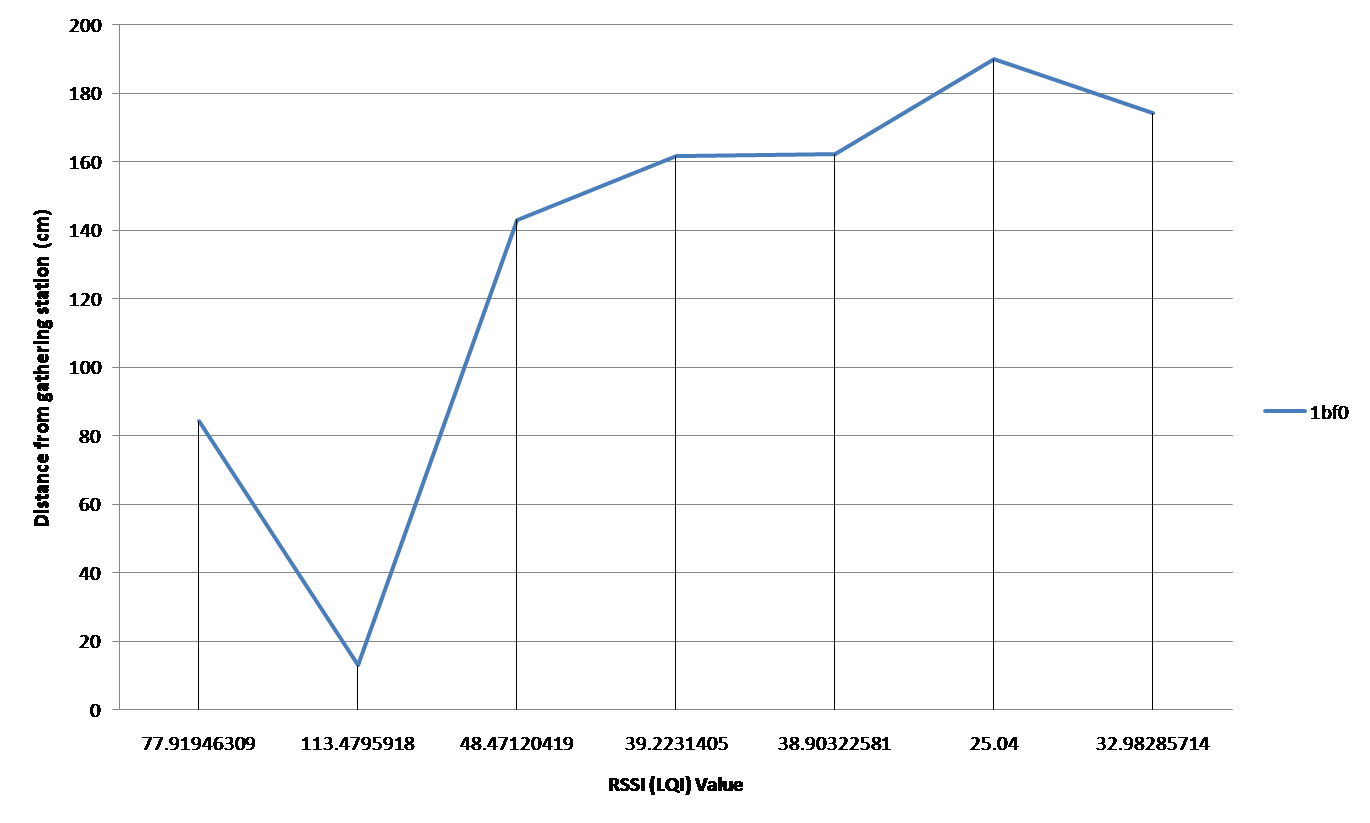}
\caption{\label{lqi4} Distance derived from experimental RSSI values}
\end{figure}

Depending on the environment (space vacuum, office with furniture, football field etc.), the user of the above formula must set $n$ accordingly. We used a value of $n=2$. In figures ~\ref{lqi1},~\ref{lqi2},~\ref{lqi3},~\ref{lqi4} we show the results from the real experiments.
From the models mentioned above, we decided to use the Random Error model to test out method, since it came closer to the real experiments data, it was simple enough to calculate and was non-deterministic.


\subsection{Clustering Methods}
\label{sec:methods}
In order to form the cluster of the intersection points, we propose 3 different methods. Before applying each method, we calculated the all the intersection points and made them available to them. 

We will explain each method in the following paragraphs:
\subsubsection{Method 1}
\begin{figure}
\centering
\includegraphics[width=0.45\textwidth]{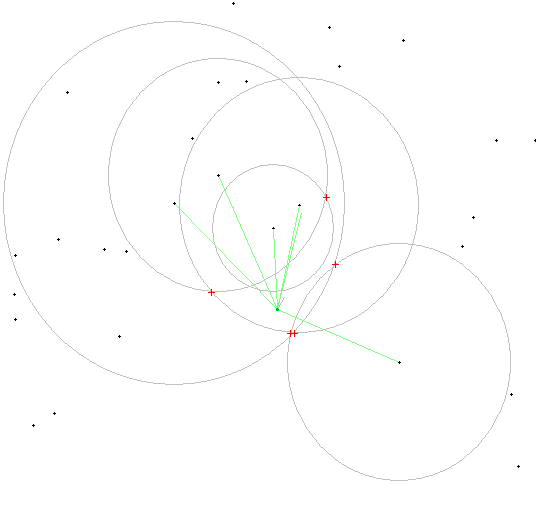}
\caption{\label{fig-4} Method 1 Cluster Example}
\end{figure}

In this method we examine the intersection points between each pair of circles. We proceed only if there are intersection points for each pair of circles. For each intersection point, we assign 0 Favour Points. We then compare the distance of each intersection point to the center of all the rest of the circles. The intersection point that is closest to each center is awarded a Favour Point. We iterate for all circles, excluding the two circles that we are examining. At the end, if the Favour Points for one intersection point are greater than zero while the Favour Points for the other intersection point are zero, we include the first intersection point in the cluster. If the Favour Points are greater than zero in both intersection points, we do not include either one in the cluster. Similarly, we include no points if both points have Favour Points equal to zero. An example of an application of this method can be seen in Figure~\ref{fig-4}. The intersection points that are included in the cluster are marked. 

\subsubsection{Method 2}
\begin{figure}
\centering
\includegraphics[width=0.45\textwidth]{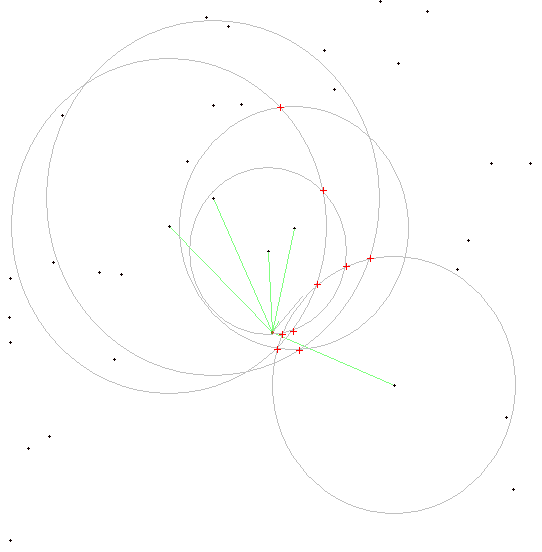}
\caption{\label{fig-5} Method 2 Cluster Example}
\end{figure}

In this method, we check to see if the intersection points are included in the rest of the circles. If an intersection point is inside all the other circles, we include this point to the cluster. A point is considered to be inside a circle if the distance between the point and the center of the circle is less than the circle radius. An example of an application of this method can be seen in Figure~\ref{fig-5}. The intersection points that are included in the cluster are marked.
\subsubsection{Method 3}
\begin{figure}
\centering
\includegraphics[width=0.45\textwidth]{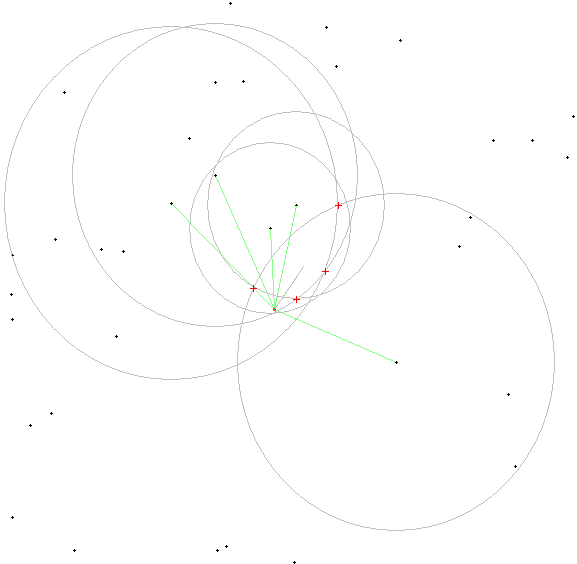}
\caption{\label{fig-6} Method 3 Cluster Example}
\end{figure}

This method is similar to the first method, but much more strict when it comes to the condition which allows an intersection point to be included in the cluster. In order for this to happen, the Favour Points of the first intersection point must be equal to the total number of circles (minus the two circles we examine in each iteration). In any other case, both intersection points are rejected. An example of an application of this method can be seen in Figure~\ref{fig-6}. The intersection points that are included in the cluster are marked.

After applying a method to form a cluster of intersection points, we calculate the final position of the node by using the coordinates of all the cluster points. We calculate the final position's X by averaging all the X coordinates of the cluster points. Simirarly, we calculate the final position's Y by averaging all the Y coordinates of the cluster points.
\subsection{Evaluation}
\label{sec:evaluation}
\subsubsection{Simulation Environment}
In order to evaluate our methods, we used the WSNGE Toolkit and Simulator \cite{wsnge} . We extended its functionality to include multilateration localization and we used its visualization system to fine tune our implementation. The nodes were uni-formally distributed in the network area and radio range $R$ was equal for all nodes. In order to obtain statistics for a wide range of estimation errors we run all three methods for 200 times, each time adding 0.001 to the error $e$ . We start the error $e$ from $0$ as the minimum and after the iterations we have an error $e$ of $0.2$. We then estimate the distance $EstDist$, by changing the real distance $RealDist$ using the following formula: $EstDist$=$RealDist$ $\pm$ $RealDist$ * $e$. The estimated distance is the radius for the circle for each neighbouring anchor.
After drawing the circles, we iterate through each pair of circles and we calculate the intersection points. We then use each of the methods described above to create a cluster and estimate the position. We calculate the error as the distance between each node's estimated position and its real position. For each iteration, we calculate the total error as : $Total Error$=$Sum(Node Errors)/(Number of Localized Nodes)$. 
We repeated the above procedure for 4 different networks, all sized $1$x$1$ with $100$ nodes. The networks had the characteristics seen in Table ~\ref{tbl-1}. Radius refers the communication radius of each node (using the Unit Disc Graph communication model) while Mean refers to the mean connectivity of the network.
Since the purpose of the above experiments was to conclude on the accuracy of our methods, we did not use localization information propagation. Instead, we assumed that all the neighbours of a single node are anchor nodes, so the mean connectivity of the network is also the average number of anchors per node.
\begin{table}[h!b!p!]
\caption{Table (Networks Used for Testing)}
\begin{tabular}{llll}
\hline
Size & Nodes & Radius & Mean \\
\hline
1$x$1 & 100 & 0.04 & 4.582 \\
1$x$1 & 100 & 0.05 & 7.199 \\
1$x$1 & 100 & 0.06 & 10.394 \\
1$x$1 & 100 & 0.07 & 13.96\\
\hline
\end{tabular}
\label{tbl-1}
\end{table}

\subsubsection{Simulation Results}

\begin{figure}
\centering
\includegraphics[width=0.45\textwidth]{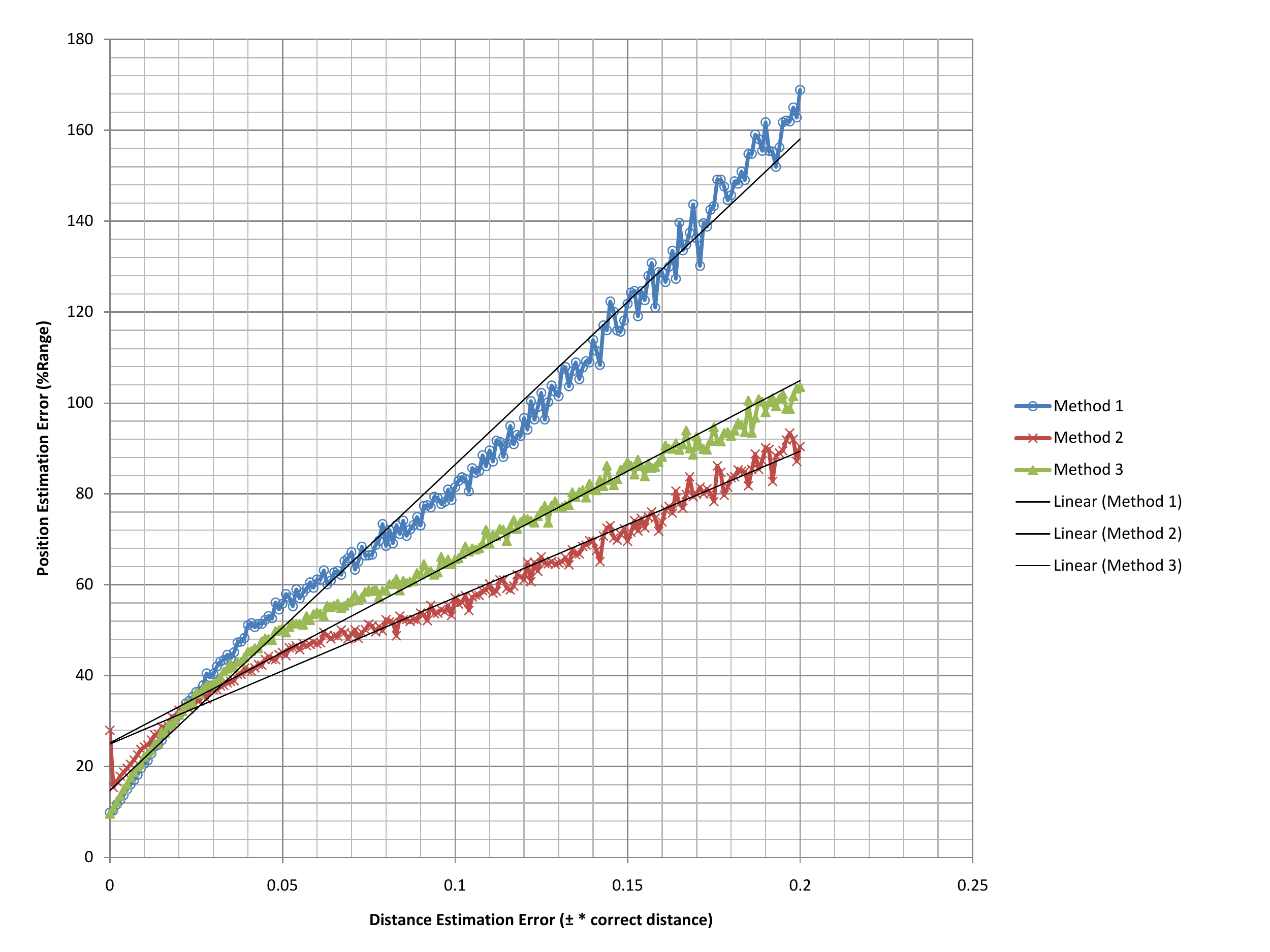}
\caption{\label{res-fig-1} Network 1 Results}
\end{figure}

For the first network, Figure~\ref{res-fig-1} shows the behaviour of the three methods for all 200 iterations.

\begin{figure}
\centering
\includegraphics[width=0.45\textwidth]{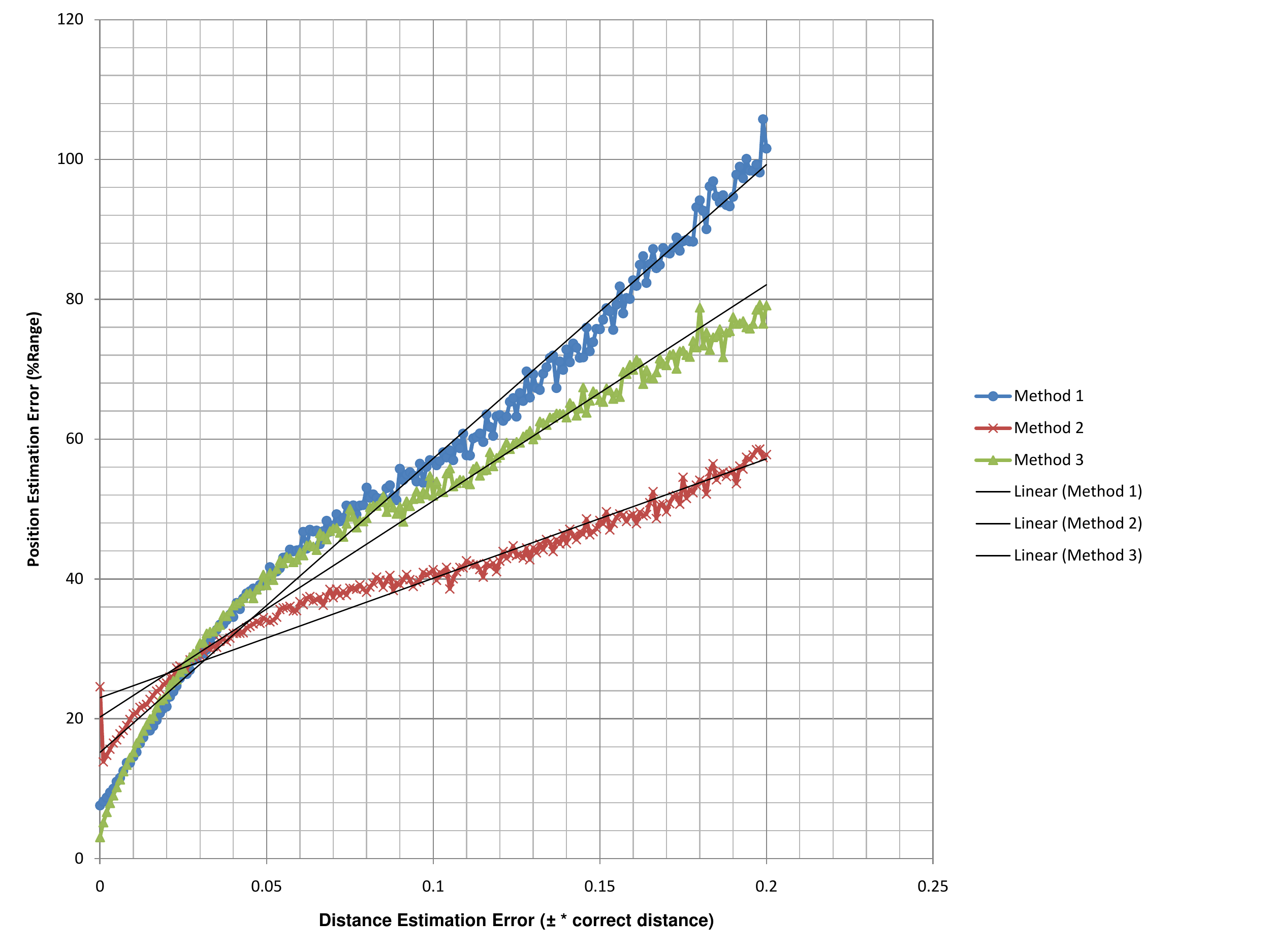}
\caption{\label{res-fig-2} Network 2 Results}
\end{figure}

For the second network,Figure~\ref{res-fig-2} shows the behaviour of the three methods for all 200 iterations.

\begin{figure}
\centering
\includegraphics[width=0.45\textwidth]{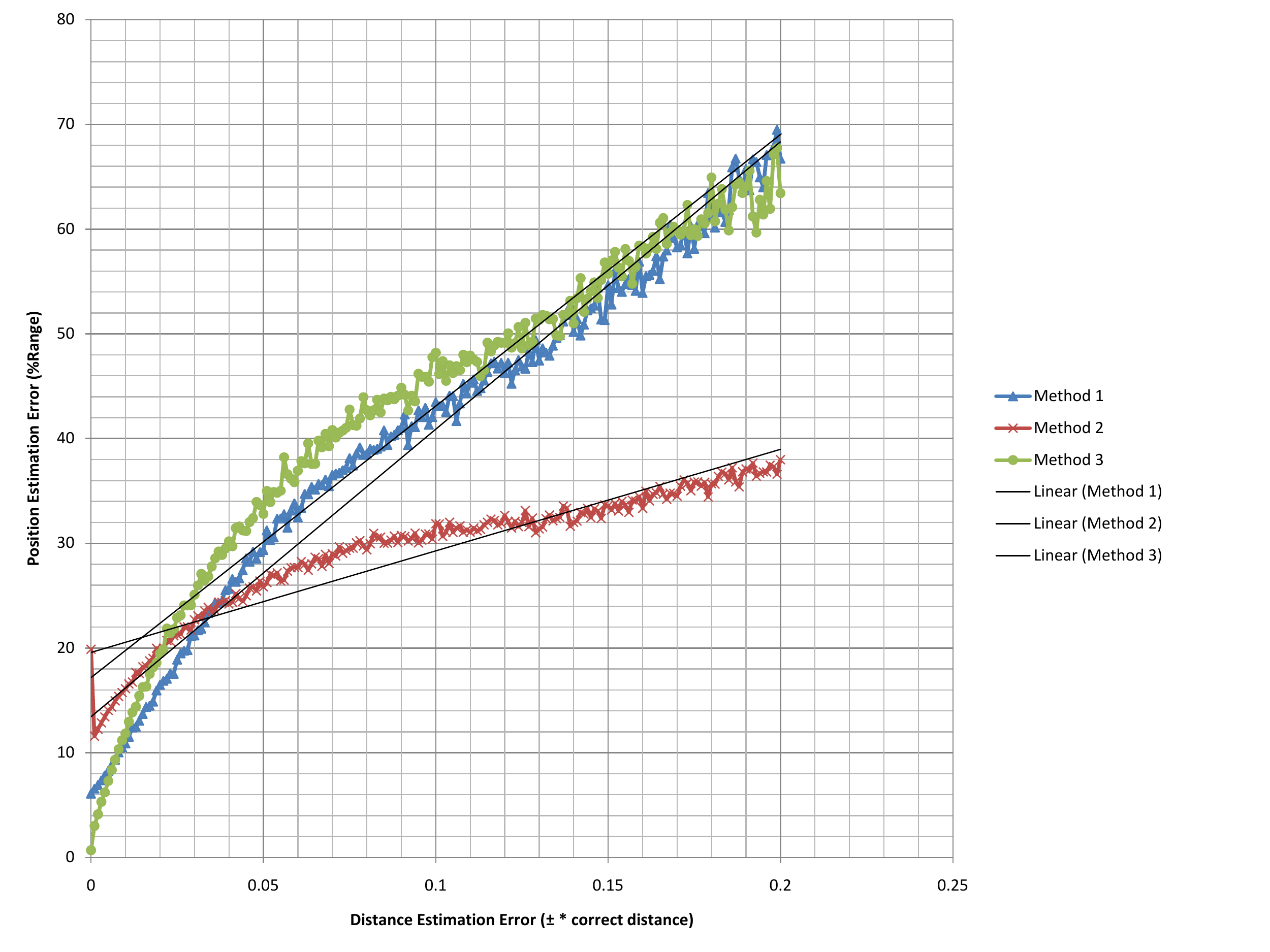}
\caption{\label{res-fig-3} Network 3 Results}
\end{figure}

For the third network, Figure~\ref{res-fig-3} shows the behaviour of the three methods for all 200 iterations.

\begin{figure}
\centering
\includegraphics[width=0.45\textwidth]{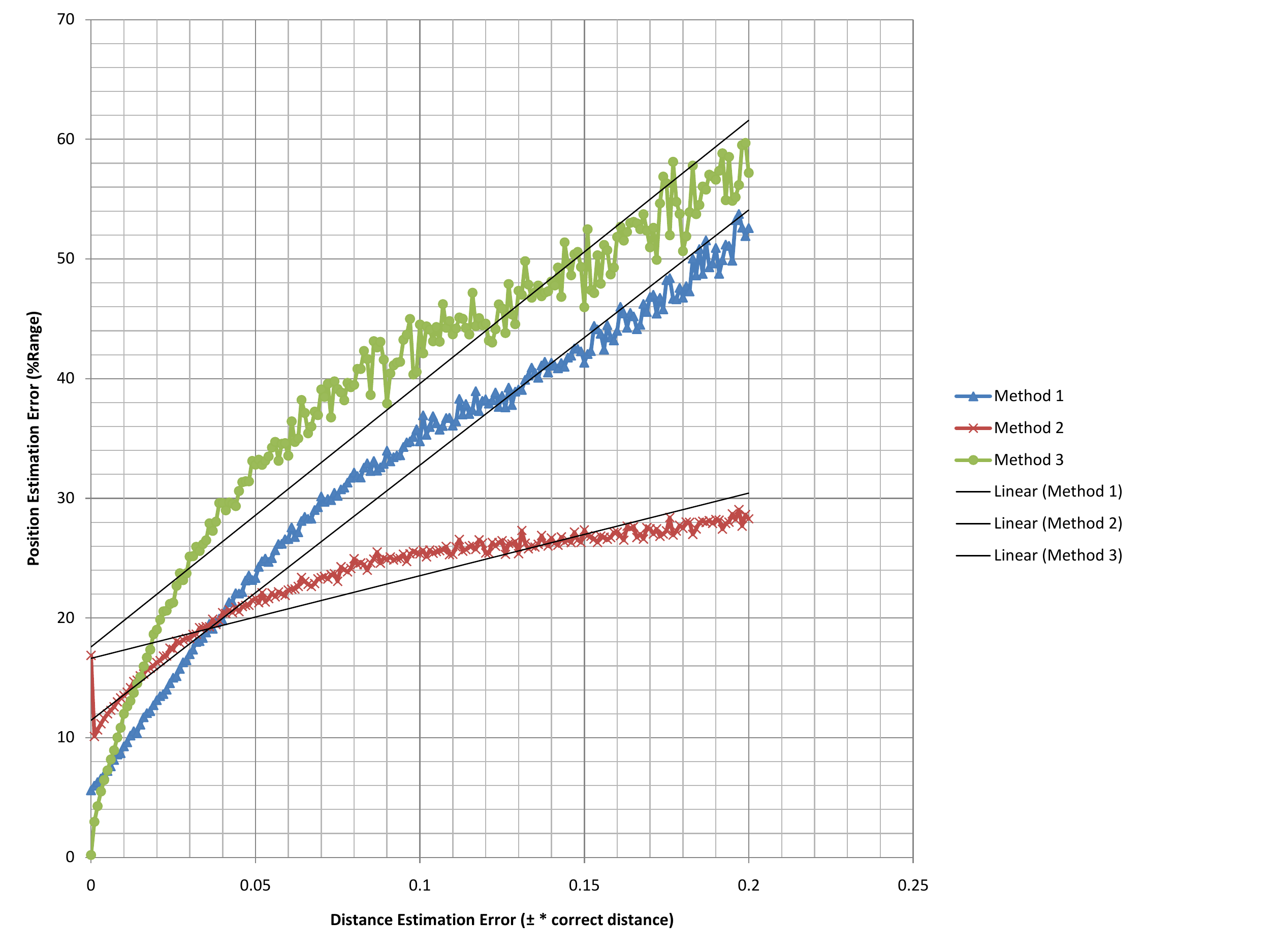}
\caption{\label{res-fig-4} Network 4 Results}
\end{figure}

For the fourth network, Figure~\ref{res-fig-4} shows the behaviour of the three methods for all 200 iterations.

The lines connect the node which is trying to find its location with its neighbours which are considered anchors for the purpose of evaluating our methods. The circles have a radius equal to the reported distance from the node to the anchor, which contains a random error. The crosses show the intersection points of the circles, which have been chosen by the respective method. 
The values on the Y axis represent the error in percentage with respect to the communication range of the nodes (\%range). In the X axis, the values represent the percentage of the error $percentageError$ with respect to the true distance from each node to an anchor. The actual distance value for each pair of node-neighbour is chosen uni-formally from $[RealDistance-RealDistance*percentageError,RealDistance+RealDistance*percentageError]$.

By observing the above results, we noticed that among our three methods, Method 2 shows the greatest robustness, since it provided consistently lower error percentages above 3$\%$ of distance estimation error in all 4 networks. It also provided us with ever decreasing slopes when increasing the mean connectivity, while the other two methods had an almost identical slope, albeit lower percentages of position estimation error when the mean connectivity increased. We also noticed that Methods 1 and 2 were not completely accurate when provided with very small distance estimation error ($<0.5\%$) while Method 3 was. In fact, Method 3 provided a $0\%$ position estimation error when the distance estimation error was also $0$, while the other 2 methods gave a higher position estimation error.

\section{Conclusions and Future Work}

In this paper, we evaluated three new methods for localizing a node using multilateration when distance estimation error exists. We used the estimated distances to draw circles and then used the intersection points of these circles to create clusters. We formed different clusters using the different methods and evaluated the resulting error in position estimation.
Future work includes exploring more parameters like different network connectivities. We can also use different communication models to evaluate our methods' robustness and accuracy.

\section {Acknowledgments}
We thank Dr. Alan Kaminsky from the Rochester Institute of Technology for the initial idea that inspired us to proceed with a more thorough investigation.

{\small
\bibliographystyle{alpha}
\bibliography{multilateration}

\newcommand{\etalchar}[1]{$^{#1}$}
\begin{thebibliography}{AEG{\etalchar{+}}06}

\bibitem[AEG{\etalchar{+}}06]{AspnesEGMWYAB2006}
J.~Aspnes, T.~Eren, D.~K. Goldenberg, A.~S. Morse, W.~Whiteley, Y.~R. Yang,
  B.~D.~O. Anderson, and P.~N. Belhumeur.
\newblock A theory of network localization.
\newblock {\em IEEE Transactions on Mobile Computing}, 5(12):1663--1678,
  December 2006.

\bibitem[AHSK05]{local3}
A.~Tarighat A.~H.~Sayed and N.~Khajehnouri.
\newblock Network-based wireless location: challenges faced in developing
  techniques for accurate wireless location informations.
\newblock {\em IEEE Signal Processing Magazine}, 22(4):24--40, 2005.

\bibitem[BHE00]{Bulusu00gps-lesslow}
Nirupama Bulusu, John Heidemann, and Deborah Estrin.
\newblock Gps-less low cost outdoor localization for very small devices, 2000.

\bibitem[CA04]{vcs2}
Qing Cao and Tarek~F. Abdelzaher.
\newblock A scalable logical coordinates framework for routing in wireless
  sensor networks.
\newblock In {\em RTSS}. ACM, 2004.

\bibitem[DMNJ04]{vcs3}
M.~E.~Goldsby D.~M.~Nicol and M.~M. Johnson.
\newblock Simulation analysis of virtual geographic routing.
\newblock In {\em Proceedings of the 2004 Winter Simulation Conference}. ACM,
  2004.

\bibitem[GG05]{local2}
F.~Gustafsson and F.~Gunnarsson.
\newblock Mobile positioning using wireless networks: possibilities and
  fundamental limitations based on available wireless network measurements.
\newblock {\em IEEE Signal Processing Magazine}, 22(4):41--53, 2005.

\bibitem[{Kar}09]{wsnge}
{Rolim Jose} {Karagiannis Marios}, {Chantzigiannakis Ioannis}.
\newblock Wsnge: A platform for simulating complex wireless sensor networks
  supporting rich network visualization and online interactivity.
\newblock In {\em Proceedings of the 42nd Annual Simulation Symposium
  (ANSS09)}, San Diego, California, USA, March 2009.

\bibitem[NN03]{Niculescu03adhoc}
Dragos Niculescu and Badri Nath.
\newblock Ad hoc positioning system (aps) using aoa, 2003.

\bibitem[ScHS01]{Savvides01dynamicfine-grained}
Andreas Savvides, Chih chieh Han, and Mani~B. Strivastava.
\newblock Dynamic fine-grained localization in ad-hoc networks of sensors,
  2001.

\bibitem[SGL05]{local1}
G.~Wei S.~Guolin, C.~Jie and K.~J.~R. Liu.
\newblock Signal processing techniques in network-aided positioning: a survey
  of state-of-the-art positioning designs.
\newblock {\em IEEE Signal Processing Magazine}, 22(4):13--23, 2005.

\bibitem[SKPP07]{1298094}
Tsenka Stoyanova, Fotis Kerasiotis, Aggeliki Prayati, and George Papadopoulos.
\newblock Evaluation of impact factors on rss accuracy for localization and
  tracking applications.
\newblock In {\em MobiWac '07: Proceedings of the 5th ACM international
  workshop on Mobility management and wireless access}, pages 9--16, New York,
  NY, USA, 2007. ACM.

\bibitem[SRL01]{Savarese01robustpositioning}
Chris Savarese, Jan Rabaey, and Koen Langendoen.
\newblock Robust positioning algorithms for distributed ad-hoc wireless sensor
  networks.
\newblock In {\em in USENIX Technical Annual Conference}, pages 317--327, 2001.

\bibitem[TM04]{vcs1}
M.~Wattenhofer R.~Wattenhofer T.~Moscibroda, R.~O'Dell.
\newblock Virtual coordinates for ad hoc and sensor networks.
\newblock In {\em ACM DIALM-POMC}. ACM, Oct 2004.

\bibitem[UR10]{TR-2010-UNIGE-RACTI-LQI}
{University of Geneva (UNIGE)} and {Research Academic Computer Technology
  Institute (CTI)}.
\newblock {LQI values heat-map with iSense hardware technical report
  (TR-2010-UNIGE-RACTI-LQI)}.
\newblock Technical report, The WISEBED consortium, 2010.
\newblock http://www.wisebed.eu.

\bibitem[{Whi}06]{1127825}
{Culler David} {Whitehouse Kamin}.
\newblock A robustness analysis of multi-hop ranging-based localization
  approximations.
\newblock In {\em IPSN '06: Proceedings of the 5th international conference on
  Information processing in sensor networks}, pages 317--325, New York, NY,
  USA, 2006. ACM.

\bibitem[WKC07]{1234829}
Kamin Whitehouse, Chris Karlof, and David Culler.
\newblock A practical evaluation of radio signal strength for ranging-based
  localization.
\newblock {\em SIGMOBILE Mob. Comput. Commun. Rev.}, 11(1):41--52, 2007.

\end{thebibliography}
}

\end{document}